\documentclass[prd,nofootinbib]{revtex4}
\usepackage{amssymb}
\usepackage{graphicx}
\usepackage{latexsym}

\begin{document}

\title{Constraining Primordial Non-Gaussianity with High-Redshift Probes}

\author{Jun-Qing Xia${}^1$}
\email{xia@sissa.it}
\author{Anna Bonaldi${}^2$}
\email{anna.bonaldi@oapd.inaf.it}
\author{Carlo Baccigalupi${}^{1,3,4}$}
\email{bacci@sissa.it}
\author{Gianfranco~De~Zotti${}^{2,1}$}
\email{gianfranco.dezotti@oapd.inaf.it}
\author{Sabino Matarrese${}^{5,6}$}
\email{sabino.matarrese@pd.infn.it}
\author{Licia Verde${}^7$}
\email{liciaverde@icc.ub.edu}
\author{Matteo Viel${}^{3,4}$}
\email{viel@oats.inaf.it}

\affiliation{${}^1$Scuola Internazionale Superiore di Studi
Avanzati, Via Bonomea 265, I-34136 Trieste, Italy}

\affiliation{${}^2$INAF-Osservatorio Astronomico di Padova, Vicolo
dell'Osservatorio 5, I-35122 Padova, Italy}

\affiliation{${}^3$INAF-Osservatorio Astronomico di Trieste, Via
G.B. Tiepolo 11, I-34131 Trieste, Italy}

\affiliation{${}^4$INFN/National Institute for Nuclear Physics, Via
Valerio 2, I-34127 Trieste, Italy}

\affiliation{${}^5$Dipartimento di Fisica ``G. Galilei",
Universit\`a di Padova, Via Marzolo 8, I-35131 Padova, Italy}

\affiliation{${}^6$INFN, Sezione di Padova, Via Marzolo 8, I-35131
Padova, Italy}

\affiliation{${}^7$ICREA (Instituci\'o Catalana de Recerca i Estudis
Avan\c{c}ats) and Instituto de Ciencias del Cosmos,(ICC-UB-IEEC)
Universidad de Barcelona, Marti i Franques 1, 08028, Barcelona,
Spain}

\date{\today}

\begin{abstract}

We present an analysis of the constraints on the amplitude of
primordial non-Gaussianity of local type described by the
dimensionless parameter $f_{\rm NL}$. These constraints are set by
the auto-correlation functions (ACFs) of two large scale structure
probes, the radio sources from NRAO VLA Sky Survey (NVSS) and the
QSO catalogue of Sloan Digital Sky Survey Release Six (SDSS DR6
QSOs), as well as by their cross-correlation functions (CCFs) with
the cosmic microwave background (CMB) temperature map (Integrated
Sachs-Wolfe effect). Several systematic effects that may affect the
observational estimates of the ACFs and of the CCFs are
investigated 
and conservatively accounted for.
Our approach exploits the large-scale scale-dependence of the
non-Gaussian halo bias. The derived constraints on {$f_{\rm NL}$}
coming from the NVSS CCF and from the QSO ACF and CCF are weaker
than those previously obtained from the NVSS ACF, but still
consistent with them. Finally, we obtain the constraints on $f_{\rm
NL}=53\pm25$ ($1\,\sigma$) and $f_{\rm NL}=58\pm24$ ($1\,\sigma$)
from NVSS data and SDSS DR6 QSO data, respectively.
\end{abstract}


\maketitle

\section{Introduction}\label{int}

The physical mechanisms responsible for the generation of primordial
perturbations seeding present-day large-scale structure,
may leave their imprint 
in the form of small deviations from a Gaussian distribution of the
primordial perturbations. Searches for primordial non-Gaussianity
can thereby provide key information on the origin and evolution of
cosmological structures (e.g., ref. \cite{Komatsuwhitepaper} and
references therein). Although the standard single-field, slow-roll,
canonical kinetic energy and adiabatic vacuum state inflation
generates very small non-Gaussianity, any inflationary model that
deviates from this may entail a larger level of it (refs.
\cite{bmr04,Komatsu2010} and references therein).

Deviations from Gaussian initial conditions are often taken to be of
the so-called local type and are parameterized by the constant
dimensionless parameter {$f_{\rm NL}$} \cite{Salopekbond90,
Ganguietal94, VWHK00, KS01, BabichCreminelliZaldarriaga}:
\begin{equation}
\Phi=\phi+f_{\rm NL}\left(\phi^2-\langle\phi^2\rangle\right)~,
\label{eq:fnl}
\end{equation}
where $\Phi$ denotes Bardeen's gauge-invariant potential and $\phi$
is a Gaussian random field. In the literature there are two
conventions: in the large scale structure (LSS) convention $\Phi$ is
linearly extrapolated to $z=0$, while in the cosmic microwave
background (CMB) convention it is evaluated deep in the matter era.
Thus, $f^{\rm LSS}_{\rm   NL} = [g(z=\infty)/g(z=0)] f^{\rm
CMB}_{\rm NL}\sim1.3f^{\rm CMB}_{\rm NL}$, where $g(z)$ denotes the
$\Lambda$-induced linear growth suppression factor. In this paper we
will use the CMB convention.


A new method \citep{DDHS08,MV08} for constraining non-Gaussianity
from large-scale structure surveys exploits the fact that the
clustering of dark matter halos --where galaxies form-- on large
scales increases (decreases) for positive (negative) $f_{\rm NL}$.
In particular, a non-Gaussianity described by eq.~(\ref{eq:fnl}),
introduces a scale-dependent boost of the halo power spectrum
proportional to $1/k^2$ on large scales ($k<0.03\,h/$Mpc), which
evolves roughly as $(1+z)$.  LSS surveys covering large volumes are
needed to access the scales where the signal arises (e.g., ref.
\cite{melita} and references therein). Among the many
currently-available tracers of the LSS, the radio sources from NRAO
VLA Sky Survey (NVSS) \cite{Condon:1998iy} and the QSO catalogue
of Sloan Digital Sky Survey Release Six (SDSS DR6 QSOs)
\cite{richards09} are particularly interesting since they span large
volumes extending out to substantial redshifts \cite{hoetal08}.
Indeed these source samples were shown to provide tight constraints
on primordial non-Gaussianity \cite{Slosar,afshorditolley,xiaACF}.

Extragalactic radio sources are uniquely well suited to probe
clustering on the largest scales. Radio surveys are in fact
unaffected by dust extinction which may introduce in the observed
large-scale distribution spurious features reflecting the
inhomogeneous extinction due to Galactic dust. Moreover, due to
their strong cosmological evolution, radio sources are very rare
locally, so that radio samples are free from the profusion of local
objects that dominate optically selected galaxy samples and tend to
swamp very large-scale structures at cosmological distances; thanks
to the strong cosmological evolution, even relatively shallow radio
surveys reach out to substantial redshifts. The NVSS
\cite{Condon:1998iy} offers the most extensive sky coverage ($82\%$
of the sky to a completeness limit of about 3 mJy at 1.4 GHz) with
sufficient statistics to allow an accurate determination of the
angular correlation function, $w(\theta)$ on scales of up to several
degrees \cite{BlakeWall,Overzier}.

In a recent paper \cite{xiaACF} it was shown that the observed NVSS
Auto-Correlation Function (ACF) hints at a positive value of
{$f_{\rm NL}$} at about the $3\,\sigma$ confidence level. This is
because the ACF is found to be still positive on angular scales
$\theta > 4^\circ$, which, for the median source redshift ($z_{\rm
m}\simeq 1$), correspond to linear scales where the correlation
function should be negative if the density fluctuation field is
Gaussian. A positive {$f_{\rm NL}$} adds power on large angular
scales, accounting for the observed ACF.

A cross-check for a positive {$f_{\rm NL}$} can be provided by an
enhancement, compared to the Gaussian case, of the Cross-Correlation
Function (CCF) of the CMB with LSS probes (late-time Integrated
Sachs-Wolfe (ISW) \cite{ISW} effect). Earlier analyses \cite{Slosar}
did not find any evidence for $f_{\rm NL}> 0$ from the NVSS-CMB CCF,
although the NVSS-CMB cross-spectrum showed an anomalous point,
suggestive of some unaccounted systematic effect in the spatial
distribution of NVSS sources and/or in the adopted CMB map. In this
paper we will revisit this issue investigating possible systematics
that may affect the results.

We will also revisit the ACF of SDSS DR6 QSOs \cite{richards09} and
their CCF. This was previously analyzed by ref. \cite{xiaisw} to
exploit the high redshift regime for early dark energy models and
for the ISW effect. Here we use up-to-date CMB maps to derive
constraints on primordial non-Gaussianity, following the approach of
ref. \cite{Slosar}.
Optically selected QSOs are also well suited to test primordial
non-Gaussianity, as they probe large-volumes and high redshifts and
are not seriously affected by dust; their major contaminating source
being stars from our own galaxy.  Ref. \cite{Slosar} used an
extension of the SDSS DR3 QSOs sample to constrain $f_{\rm NL}$.
Here we revisit the analysis using an improved bias model, improved
QSO catalog, updated knowledge of the sources redshift distribution,
up-to-date CMB maps and complementary analysis methods.


The structure of the paper is as follows. In \S\,\ref{theory} we
review the effects of primordial non-Gaussianity on the ACF and on
the CCF, and the theory of the late-time ISW effect.
Section~\ref{data} contains the analysis of ACF and CCF for NVSS
radio sources and SDSS DR6 QSOs. In \S\,\ref{method} we present the
method used to derive constraints on $f_{\rm NL}$.
Section~\ref{result} contains our main results. We conclude with a
discussion and comparison with related work  in \S\,\ref{summary}.

\section{Theoretical Framework}\label{theory}

In this section we briefly present the equations describing the ISW
effect, and review the impact of primordial non-Gaussianity on
cosmic observables relevant to our analysis, namely the halo mass
function and the halo bias.

\subsection{Integrated Sachs-Wolfe effect}

The temperature anisotropy due to the ISW effect can be expressed as
an integral of the time derivative of the gravitational potential
$\Phi$ over conformal time $\eta$
\begin{equation}
{\frac{\Delta T}{T}}^{\rm ISW}(\hat{\bf n})=-2\int{\dot{\Phi}[\eta,\hat{\bf
n}(\eta_0-\eta)]d\eta}~.\label{isweq}
\end{equation}
A CMB photon falling into a gravitational potential well gains
energy, while loses energy when it climbs out of it. These effects
exactly cancel out if the potential $\Phi$ is time independent, such
as in the matter dominated era, when the fractional density contrast
$\delta_{\rm m}$ is proportional to the scale factor $a$, so that
$\dot\Phi=0$ and no ISW is produced. However, when dark energy (or
curvature) becomes important at later times, the potential evolves
as the photon passes through it and $\dot\Phi\neq0$, producing
additional CMB anisotropies.

The late-time ISW effect can thereby be a powerful tool for probing
 dark energy and its evolution. However, the main contribution of
the ISW effect to CMB anisotropies occurs on large scales that are
strongly affected by cosmic variance. This problem can be overcome
by  cross-correlating the CMB temperature fluctuations with the
distribution of extragalactic sources.

The number density contrast in a given direction $\hat{\bf n}_1$ is:
\begin{eqnarray}
\delta_{\rm g}(\hat{\bf n}_1)&=&\int{f(z)\delta_{\rm m}(\hat{\bf
n}_1,z)dz}=\int{b_{\rm g}(z)\frac{dN}{dz}(z)\delta_{\rm m}(\hat{\bf
n}_1,z)dz}~,
\end{eqnarray}
where $b_{\rm g}(z)$ is the scale-independent bias factor relating
the density contrast of visible objects, $\delta_{\rm g}$, to the
mass density contrast $\delta_{\rm m}$ ($\delta_{\rm g}=b_{\rm
g}\delta_{\rm m}$), and $dN/dz$ is the normalized redshift
distribution of the survey.

The ISW temperature anisotropy in a direction  $\hat{\bf
n}_2$ is:
\begin{equation}
\frac{\Delta T}{T}(\hat{\bf n}_2)=-2\int{\frac{d\Phi}{dz}(\hat{\bf
n}_2,z)dz}~.
\end{equation}
The ACF of a complete sample of sources, $C^{\rm gg}(\theta)$, and
its CCF with a CMB map, $C^{\rm gT}(\theta)$, as a function of the
angular separation $\theta$ between $\hat{\bf n}_1$ and $\hat{\bf
n}_2$, can be written as:
\begin{eqnarray}
C^{\rm gg}(\theta)&\equiv&\left\langle\delta_{\rm g}(\hat{\bf
n}_1)\delta_{\rm g}(\hat{\bf
n}_2)\right\rangle=\sum^{\infty}_{\ell=2}\frac{2\ell+1}{4\pi}C^{\rm
gg}_\ell P_\ell[\cos(\theta)]~,\\
C^{\rm gT}(\theta)&\equiv&\left\langle\frac{\Delta{T}}{T}(\hat{\bf
n}_1)\delta_{\rm g}(\hat{\bf
n}_2)\right\rangle=\sum^{\infty}_{\ell=2}\frac{2\ell+1}{4\pi}C^{\rm
gT}_\ell P_\ell[\cos(\theta)]\exp[-0.5(\theta_{\rm s}\ell)^2]~,
\end{eqnarray}
where the $P_\ell[\cos(\theta)]$ are the Legendre polynomials,
$\theta_{\rm s}$ is the smoothing scale of the CMB map. Here $C^{\rm
gg}_\ell$ and $C^{\rm gT}_\ell$ are the auto-correlation and
cross-correlation power spectra, respectively, given by:
\begin{eqnarray}
C^{\rm gg}_\ell&=&\frac{2}{\pi}\int{k^2dkP(k)[I^{\rm g}_\ell(k)]^2}~, \\
C^{\rm gT}_\ell&=&\frac{2}{\pi}\int{k^2dkP(k)I^{\rm
ISW}_\ell(k)I^{\rm g}_\ell(k)}~,
\end{eqnarray}
in terms of the present day matter power spectrum, $P(k)$, and of
the functions $I^{\rm g}_\ell(k)$ and $I^{\rm ISW}_\ell(k)$:
\begin{eqnarray}
I^{\rm g}_\ell(k)&=&\int{b_{\rm g}(z)\frac{dN}{dz}(z)\delta_{\rm
m}(k,z)j_\ell[k\chi(z)]dz}~, \\
I^{\rm
ISW}_\ell(k)&=&-2\int{\frac{d\Phi(k)}{dz}j_\ell[k\chi(z)]dz}~,
\label{jl}
\end{eqnarray}
$j_\ell(x)$ being the spherical Bessel functions, and $\chi$ the
comoving distance. We use the publicly available package {\tt
CAMB$_{-}$sources}\footnote{Available at http://camb.info/sources/.}
\cite{camb} to calculate the theoretical ACF and CCF.

\begin{figure*}[htpb]
\begin{center}
\includegraphics[scale=0.7]{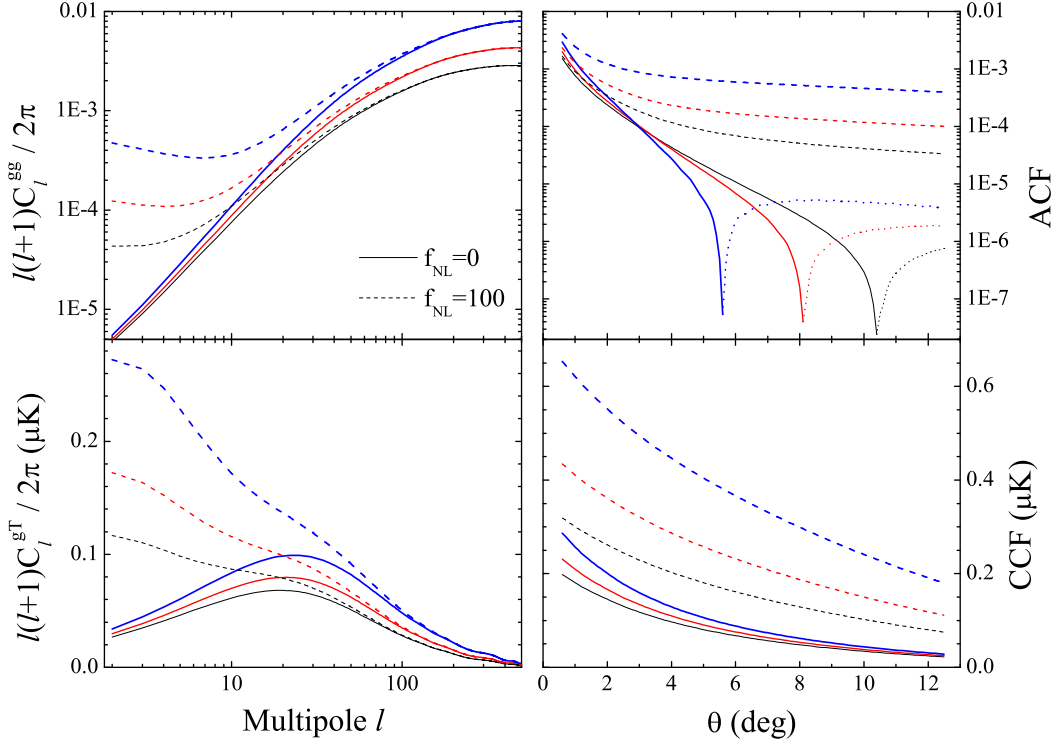}
\caption{Effects of non-Gaussianity on the auto-correlation and
cross-correlation power spectra (left panels), and on the ACF and
CCF (right panels) for three different Gaussian halo bias models:
$b=1.7$ (bottom black curves), $b=1.1+0.6/D(z)$ (middle red curves),
$b=1.1+0.6/D^2(z)$ (top blue curves). Solid curves: $f_{\rm NL}=0$;
dashed curves: $f_{\rm NL}=100$.} \label{fig:NGeffect}
\end{center}
\end{figure*}

\subsection{Effects of Primordial Non-Gaussianity}\label{sect:NG}

In the presence of primordial non-Gaussianity, the mass function
$n_{\rm NG}(M,z,f_{\rm NL})$ can be written in terms of the Gaussian
one $n_{\rm G}^{\rm sim}(M,z)$, for which a good fit to the results
of simulations is provided by e.g., the Sheth-Tormen formula
\cite{shethtormen}, multiplied by a non-Gaussian correction factor
\cite{mvj,VJKM01,loverdeetal08} \footnote{Although attempts have
been made to derive directly an expression for the non-Gaussian mass
function \cite{MR10,D'Amico:2010ta}.}:
\begin{eqnarray}
R_{\rm NG}(M,z,f_{\rm NL})=1+\frac{\sigma^2_{\rm M}}{6\delta_{\rm
ec}(z)}\left[\!S_{\rm 3,M}\!\left(\!\frac{\delta^4_{\rm
ec}(z)}{\sigma^4_{\rm M}} - 2\frac{\delta^2_{\rm
ec}(z)}{\sigma^2_{\rm M}} - 1\!\!\right)\! +\!\frac{dS_{\rm
3,M}}{d\!\ln\!{\sigma_{\rm M}}}\!\left(\!\frac{\delta^2_{\rm
ec}(z)}{\sigma^2_{\rm M}}- 1\!\!\right)\!\right]~,
\end{eqnarray}
where the normalized skewness of the density field, $S_{\rm 3,M}$,
is $\propto f_{\rm NL}$, $\sigma_{\rm M}$ denotes the rms of the
dark matter density field linearly extrapolated to $z=0$ and
smoothed on scale $R$ corresponding to the Lagrangian radius of a
halo of mass $M$, and $\delta_{\rm ec}$ is the critical density for
ellipsoidal collapse, calibrated on N-body simulations
\cite{grossi09}. For high peaks ($\delta_{\rm ec}/\sigma_{\rm M}\gg
1$) and small $f_{\rm NL}$, $\delta_{\rm ec}$ is slightly smaller
than the critical density for spherical collapse, $\delta_{\rm c}(z)
= \Delta_{\rm c}(z)D(0)/D(z)$ where $D(z)$ is the linear growth
factor, and $\Delta_{\rm c}(z)\sim 1.68$ evolves very weakly with
redshift.

The large-scale halo bias is also modified by the presence of
non-Gaussianity \citep{DDHS08,MV08,grossi09}:
\begin{equation}
b_{\rm NG}(z)-b_{\rm G}(z)\simeq2(b_{\rm G}(z)-1)f_{\rm
NL}\delta_{\rm ec}(z)\alpha_{\rm M}(k)~, \label{eq:nghalobias}
\end{equation}
where the factor $\alpha_{\rm M}(k)$ encloses the scale and halo
mass dependence. In practice, we find that, on large scales,
$\alpha_{\rm M}(k)\propto 1/k^2$, is independent of the halo mass.

We assume that the large-scale linear halo bias for the Gaussian
case is \cite{shethtormen}:
\begin{eqnarray}\label{eq:bST}
b_{\rm G}&=&1+\frac{1}{D(z_{\rm o})}\left[\frac{q\delta_{\rm
c}(z_{\rm f})}{\sigma^2_{\rm M}}-\frac{1}{\delta_{\rm c}(z_{\rm
f})}\right]+\frac{2p}{\delta_{\rm c}(z_{\rm f})D(z_{\rm
o})}\left\{1+\left[\frac{q\delta^2_{\rm c}(z_{\rm f})}{\sigma^2_{\rm
M}}\right]^p\right\}^{-1}~,
\end{eqnarray}
where $z_{\rm f}$ is the halo formation redshift, and $z_{\rm o}$ is
the halo observation redshift. As we are interested in massive
haloes, we expect that $z_{\rm f}\simeq z_{\rm o}$. Here, $q=0.75$
and $p=0.3$ account for the non-spherical collapse and are a fit to
numerical simulations (see also refs.
\cite{Mo96,Mo97,Scoccimarro01}).

Finally, the weighted effective halo bias is given by
\begin{equation}
b_{\rm NG}^{\rm eff}(M_{\rm min},z,k,f_{\rm
NL})=\frac{\int^\infty_{M_{\rm min}}b_{\rm NG}n_{\rm
NG}dM}{\int^\infty_{M_{\rm min}}n_{\rm NG}dM}~,
\end{equation}
$M_{\rm min}$ being the minimum halo mass hosting a source of the
kind we are considering.

Two things should be stressed when using eq.~(\ref{eq:nghalobias}):
a degeneracy between $b_{\rm G}$ and $f_{\rm NL}$ is expected at a
given scale (the same amount of non-Gaussian bias can be given by
different pairs of $b_{\rm G}$ and $f_{\rm NL}$ values; strictly
speaking $b_{\rm G}$ is not a free parameter here, and the
degeneracy is between $M_{\rm min}$, which is a free parameter, and
$f_{\rm NL}$; however $b_{\rm G}$ is strongly dependent on $M_{\rm
min}$); the $1/k^2$ scale-dependence implies that large-scales are
primarily affected by $f_{\rm NL}$, while small scales are mainly
affected by $b_{\rm G}$ (see a quantitative discussion for this in
ref. \cite{xiaACF}).

From the two left-hand panels of figure~\ref{fig:NGeffect}, one can
see that a positive $f_{\rm NL}$ enhances the amplitudes of the
auto-correlation and cross-correlation power spectra especially on
large angular scales ($\ell<100$). Consequently, the amplitudes of
the ACF and CCF are also enhanced (right-hand panels of figure
\ref{fig:NGeffect}). The ACF remains positive up to angular scales
$\theta > 4^\circ$ where, for the adopted redshift distribution and
a redshift-independent $M_{\rm min}$, the ACF is expected to become
negative.  The lower left-hand panel of figure~\ref{fig:NGeffect}
shows that the effects of non-Gaussianity on the cross-correlation
power spectrum become very large for the lowest multipoles.
Actually, most of the non-Gaussian contribution to the CCF comes
from $\ell<10$, corresponding to angular scales $\gtrsim 18^\circ$.
Its detection therefore requires uniform sky coverage up to very
large scales, but this is very difficult to achieve observationally.
For example, in the case of NVSS sources, very small modulations of
the source surface density on very large scales, such as those of
interest here, are easily swamped by systematic variations of the
survey effective depth correlated with the varying observing
conditions \cite{BlakeWall}. 
Also, the ``integral constraint'' \cite{WallJenkins} needs to be
dealt carefully. In fact, the ACF and CCF estimators involve
differences between local values of some quantity and their mean
value over the considered area (see eqs. (\ref{acfest}) and
(\ref{ccfest})). But if such area is not much larger than the
correlation scale, the computed mean may differ from the true mean
by a constant offset $c$, not known a priori, that must therefore be
considered as a free parameter.

\section{Observed ACF \& CCF Data}\label{data}

In this section we describe two LSS probes, NVSS radio sources and
high redshift SDSS DR6 QSOs, and present observational
determinations of their ACF and CCF with the CMB.

\subsection{NVSS Radio Sources}\label{ACFdata}
We start by confining our analysis to NVSS sources brighter than 10
mJy, since the surface density distribution of fainter sources
suffers from declination-dependent fluctuations \cite{BlakeWall}.
Density gradients in the NVSS catalog become increasingly
unimportant as the source brightness threshold is increased (see
refs. \cite{BlakeWall,CHM10}). Also we mask the strip
$|b|<5^\circ$, where the catalog may be substantially affected by
Galactic emissions. The NVSS source surface density at this
threshold is $16.9\,{\rm deg}^{-2}$ and the redshift distribution at
this flux limit has been recently determined by ref. \cite{Brookes}.
Their sample, complete to a flux density of 7.2 mJy, comprises 110
sources with $S_{1.4\rm GHz}\ge 10\,$mJy, of which 78 (71\%) have
spectroscopic redshifts, 23 have redshift estimates via the $K$--$z$
relation for radio sources, and 9 were not detected in the $K$ band
and therefore have a lower limit to $z$. We adopt here the smooth
description of this redshift distribution given by ref.
\cite{DeZotti10}:
\begin{equation}\label{eq:zdis}
dN/dz=1.29+32.37z-32.89z^2+11.13z^3-1.25z^4~.
\end{equation}
The mean redshift of this sample is $\langle z\rangle=1.23$. From
the NVSS catalog we construct a pixelized map using the HEALPix
software package \cite{healpix} with resolution $N_{\rm side} = 64$,
yielding pixel areas of $0.92^\circ\times 0.92^\circ$.

Our ACF estimator $\hat{w}(\theta)$ reads:
\begin{equation}
\hat{w}(\theta)=\frac{1}{N_{\theta}}\sum_{\rm i,j}\frac{(n_{\rm
i}-f_{\rm i}\bar{n})(n_{\rm j}-f_{\rm j}\bar{n})}{\bar{n}^2}
~\label{acfest},
\end{equation}
where $n_{\rm i}$ is the number of sources in the i-th pixel,
$f_{\rm i}$ is the un-masked fraction of the pixel area, $\bar{n}$
is the expectation value for the number of objects in the pixel area
\cite{xiaisw}. The sum runs over all pixel pairs whose centers are
separated by an angle $\theta$. The equal weighting used here is
nearly-optimal because of the uniform NVSS sky coverage and because
on large scales the noise is dominated by sample variance. We use
$N_{\rm b} = 9$ angular bins, $1^\circ$ wide. The first one is
centered at $\theta=40'$ (on smaller angular scales the dominant
contribution to the angular correlation function comes from multiple
components of individual radio galaxies); the other 8 are centered
at $1^\circ \cdots 8^\circ$ (on still larger scales the signal may
be affected or even dominated by spurious density gradients
\cite{BlakeWall,Overzier}).

The estimated NVSS ACF was previously found to be positive over the
full range of angular scales we consider \cite{xiaACF} (see also
ref. \cite{BlakeWall}), although the integral of $w(\theta)$ over
the full survey solid angle vanishes by construction. As pointed out
by ref. \cite{Negrello}, for a Gaussian distribution of primordial
density fluctuations and a realistic redshift distribution for NVSS
sources, the $w(\theta)$ must vanish at $\theta\simeq 4^\circ$ and
become negative on larger scales unless the redshift dependence of
the bias parameter for radio sources is drastically different from
that of optical QSOs. Consistency with the clustering properties of
the latter sources can be recovered allowing for a small amount of
non-Gaussianity ($f_{\rm NL}=62 \pm 27$ ($1\,\sigma$)
\cite{xiaACF}).

Since the NVSS covers most of the sky it can safely be assumed to
provide a ``fair sample'' of the universe, so that the derived
angular correlation function is not affected by the ``integral
constraint'' significantly. Nevertheless, in ref. \cite{xiaACF} we
have explored whether the indications of an excess (compared to the
Gaussian case) positive contribution to the ACF on large scales may
be due to a unexpectedly large {\bf positive} offset $c$ (a negative
$c$ would obviously exacerbate the discrepancy with expectations
from the Gaussian case). To this end we have added to $w(\theta)$ a
constant $c$ and have marginalized over it, allowing this quantity
to vary in the range [$10^{-8}$, $10^{-4}$], where the lower limit
means essentially zero (we work on a logarithmic scale) while the
upper limit corresponds to the case where the offset accounts for
the full clustering signal on the largest scales (see ref.
\cite{Negrello}). The best-fit value of $c$ is about $\simeq
10^{-5}$, i.e. negligibly small, confirming that the data do not
show indications that $w(\theta)$ is appreciably affected by the
integral constraint. On the other hand, $c$ is weakly constrained by
the data, and values much larger than the adopted upper limit are
formally allowed. Such large values of $c$ would imply that $w_{\rm
true}(\theta)=w_{\rm estimated}(\theta) -c$ is negative on scales
$\theta > 4^\circ$, and the indications of non-Gaussianity would no
longer be statistically significant. Such values however are not
physically plausible because $c$ is necessarily negligibly small
when, as in the NVSS case, the survey area is so much larger than
any other relevant scale (see Sec. 9.4.2 of ref.
\cite{WallJenkins}), even in the non-Gaussian case. Allowing also
for the correction proposed by ref. \cite{ws09} to account for the
infrared divergence of the non-Gaussian halo correlation function
the constraints on {$f_{\rm NL}$}, become $f_{\rm NL}=58 \pm 28$
($1\,\sigma$) \cite{xiaACF}.

A potentially trickier issue are large-scale surface density
gradients due to instrumental effects that may spuriously enhance
the ACF estimates. This issue has been extensively discussed by ref.
\cite{BlakeWall} (see also ref. \cite{Overzier}) who concluded that
this effect is negligible if we restrict ourselves to a flux limit
of 10 mJy, as we did. An upper limit to the magnitude of the effect
is an offset of $10^{-4}$, that, as noted above, would account for
the estimated clustering signal on the largest scales. Having
allowed $c$ to take on this value, we have automatically allowed for
the corresponding  decrease of the statistical significance of
$f_{\rm NL} > 0$. As a further test of the possibility that the
large-scale clustering is spuriously enhanced by surface density
gradients due to instrumental effects we have redone the analysis
restricting ourselves to sources above 20 mJy, for which such
effects are negligible on all scales, so that $c=0$. In spite of the
poorer statistics we find indications of $f_{\rm NL} > 0$ at
approximately the same significance level as for the $S\geq10\,$mJy
sample: $f_{\rm NL}=73 \pm 32$ ($1\,\sigma$ error) and $19 < f_{\rm
NL} < 139$ (95\% confidence level).


To measure the CCF between the NVSS number density map and the CMB
map, we use the following estimator:
\begin{equation}
\hat{c}^{\rm gT}(\theta)=\frac{1}{N_{\theta}} \sum_{\rm i,j}(T_{\rm
i}-\bar{T})\frac{n_{\rm j}-f_{\rm
j}\bar{n}}{\bar{n}}~,\label{ccfest}
\end{equation}
where $T_{\rm i}$ is the CMB temperature in the $i$-th pixel and
$\bar{T}$ is the mean (monopole) value for the CMB temperature in
the unmasked area. The results presented here rely on the 7 years
Internal Linear Combination (ILC) map of the CMB provided by the
Wilkinson Microwave Anisotropy Probe (WMAP) team \citep{ilc7}. We
have checked that the using other CMB maps (the 5 years ILC map by
the WMAP team and the improved (cleaner) map by ref.
\cite{Delabrouille09}) does not change the results in any
significant way. We adopt the WMAP KQ75 mask, excluding about 30\%
of the sky at low Galactic latitude, to avoid most of the residual
Galactic contamination. This mask is then combined with the NVSS sky
coverage; regions of the NVSS catalogue that may introduce spurious
features because of missing snapshot observations and over-dense
regions associated with bright of extended sources (Table~1 of ref.
\cite{Overzier}) are also masked.

As shown in \S\,\ref{sect:NG}, the effect on the CCF of primordial
non-Gaussianity shows up primarily on very large angular scales, at
variance with the ACF for which the non-Gaussian signal is localized
at $\theta=2^\circ-5^\circ$, as shown by ref. \cite{xiaACF}. Thus
the residual large-scale systematic fluctuations in the NVSS source
density with declination due to the varying projection of the beam
and the change in observing configuration, that ref.
\cite{BlakeWall} found to be still present at the 10 mJy flux limit,
may seriously bias the CCF estimate. While it is true that these
fluctuations should be, on average, uncorrelated with the CMB signal
and therefore, on average, should not contribute to the CCF signal,
we  deal here with a single realization of the Universe, and few
independent modes; in practice, the jackknife procedure we employ to
estimate the errors is not suited to deal properly with such an
effect. We therefore restrict ourselves to $S_{1.4\rm GHz}\ge
20\,$mJy, where such fluctuations are negligible. The normalized
redshift distribution for the sample of ref. \cite{Brookes} does not
change significantly if the flux limit is increased to 20 mJy, and
therefore can still be described by eq.~(\ref{eq:zdis}). The ACF for
this sample does not differ in any significant way from that of the
$S_{1.4\rm GHz}\ge 10\,$mJy on the scales of interest.

\begin{figure}
\begin{center}
\includegraphics[scale=0.45]{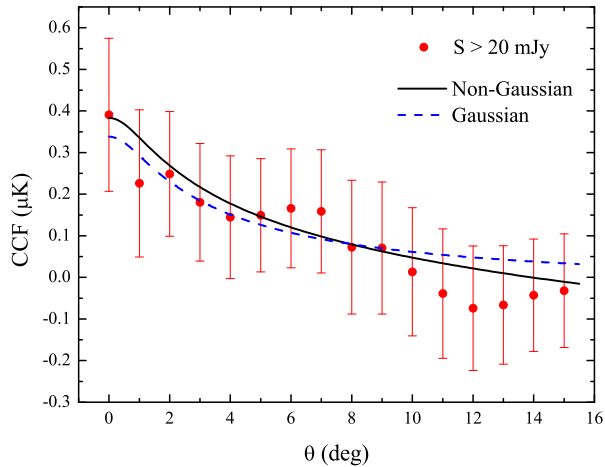}
\caption{Estimated CCF between NVSS radio sources brighter than 20
mJy and the WMAP7 ILC map. Error bars are jackknife estimated. The
black solid line is the best fit non-Gaussian model, while the blue
dashed line refers to the Gaussian case. In the latter case, $A_{\rm
amp}$ is a free parameter and the fit is obtained for $A_{\rm
amp}=1.14$. It is indeed very close to the best-fit non-Gaussian
curve, but there is no physical justification for $A_{\rm amp}\neq
1$. \label{fig:nvssccf}}
\end{center}
\end{figure}

The CCF is computed in $N_b=16$ angular bins, spaced by $1^\circ$,
with $0^\circ\leq\theta\leq15^\circ$. The covariance matrix of the
data points is estimated using the jackknife re-sampling method
\cite{covariance}, dividing the unmasked area into $M=30$ patches.
From the 30 ACF and the CCF estimates obtained neglecting a patch in
turn, we compute the diagonal (variance) and off-diagonal
(covariance) elements of the covariance matrix. Our estimate of the
NVSS-CMB CCF, shown in figure \ref{fig:nvssccf}, is fully consistent
with previous estimates using different approaches
\cite{gianna08,hoetal08,Raccanelli,Massardi10,CHM10}.

\subsection{SDSS DR6 QSOs}

The SDSS DR6 QSO catalog released by ref. \cite{richards09} contains
about $10^6$ objects with photometric redshifts ranging from $0.065$
to $6.075$ over a total area of $8417\,{\rm deg}^2$ ($\sim20\%$ of
the whole sky). We refer the reader to ref. \cite{richards09} for a
detailed description of the object selection with the non-parametric
Bayesian classification kernel density estimator (NBC-KDE)
algorithm. We use the electronically-published table that contains
only objects with the ``good'' flag with values in the range
$[0,6]$. The higher the value, the more probable for the object to
be a real QSO (see \S\,4.2 of ref. \cite{richards09} for details).
Furthermore we restrict ourselves to the ``uvxts=1'', i.e. to QSOs
clearly showing a UV excess which should be a signature of a QSO
spectrum (in this case we have $N_{\rm qso}\approx6\times10^5$
QSOs). In order to minimize the effect of Galactic extinction on the
observed QSO distribution, we  mask regions with $A_{\rm g}\ge
0.18$.

We fit the redshift distribution $dN/dz$ of the DR6 QSO sample with
a function of the form:
\begin{equation}
\frac{dN}{dz}(z)=\frac{\beta}{\Gamma(\frac{m+1}{\beta})}\frac{z^m}{z^{m+1}_0}
\exp\left[-\left(\frac{z}{z_0}\right)^\beta\right]~.\label{reddis}
\end{equation}
The best-fit values of the parameters are $m=2.00$, $\beta=2.20$,
$z_0=1.62$; the mean redshift of the sample is $\bar{z}\sim1.49$.

\begin{figure*}
\begin{center}
\includegraphics[scale=0.43]{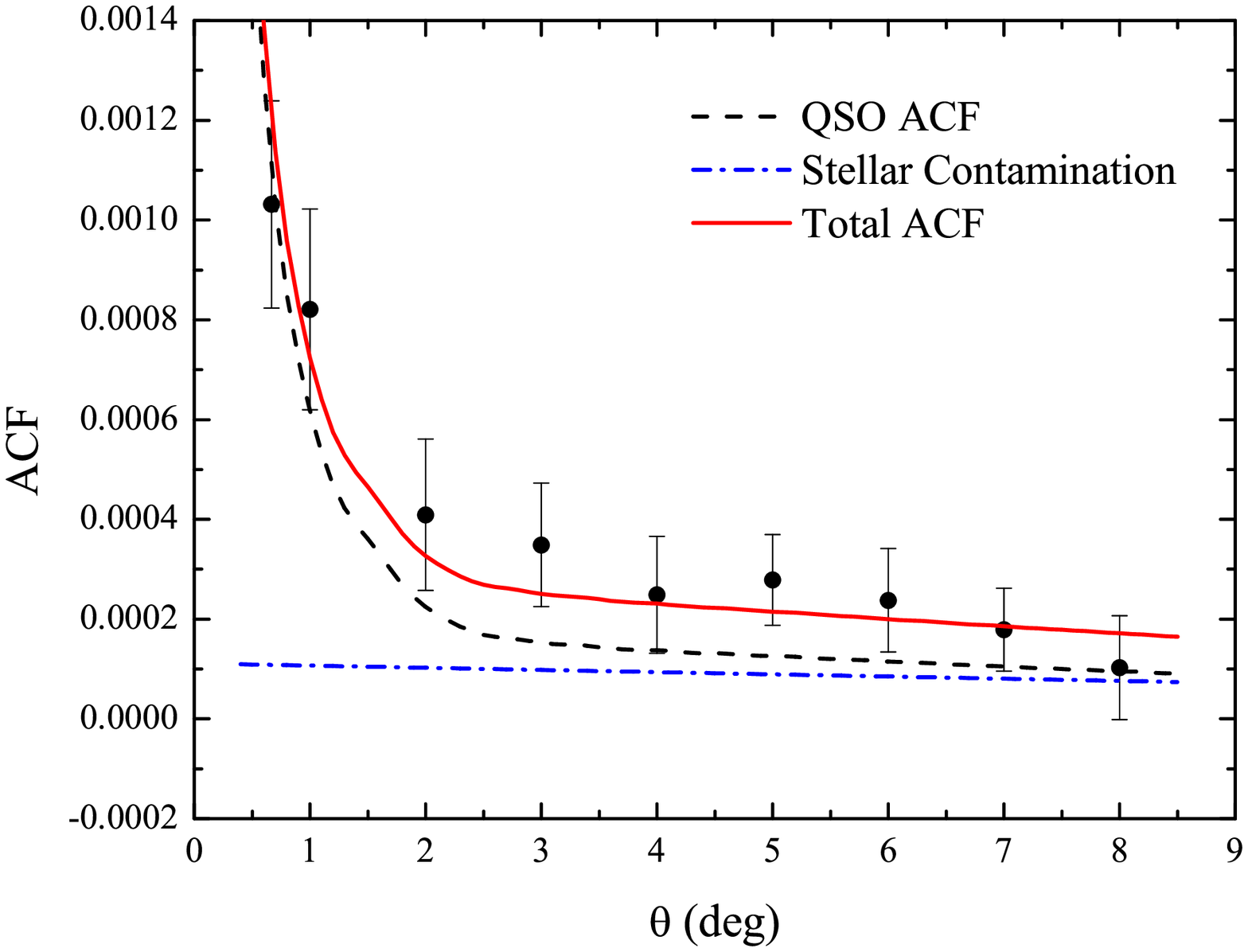}
\includegraphics[scale=0.43]{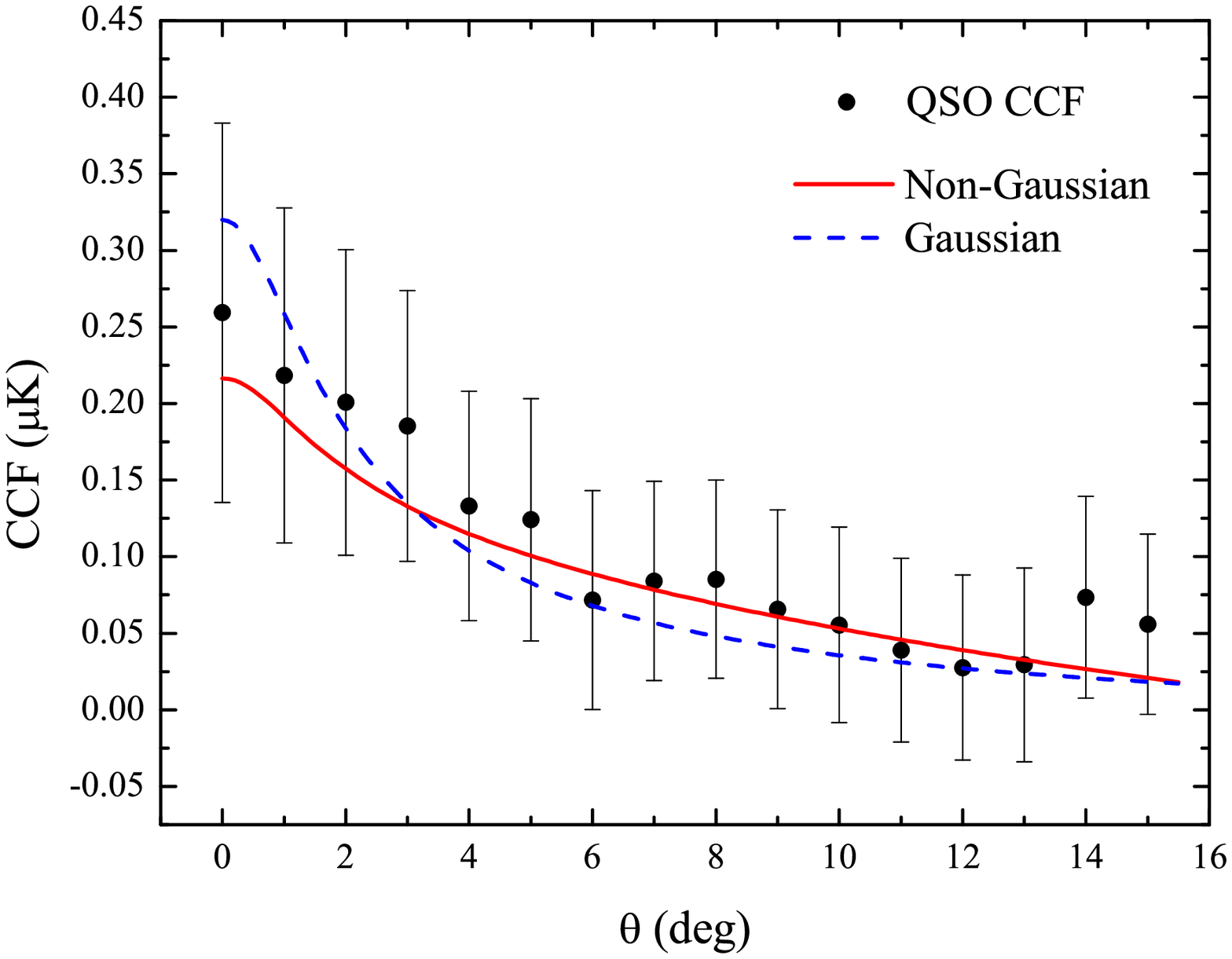}
\caption{Estimated ACF of SDSS DR6 QSOs and CCF between their
distribution and the WMAP7 ILC map. Error bars are jackknife
estimated. We also show the predictions from our best fit model (red
solid lines). In the left panel the black dashed line and the blue
dash-dotted line show the estimated contributions from QSOs and from
stars contaminating the sample to the global model ACF (red solid
line). In the right panel we also show the the best fit model for
the Gaussian case (blue dashed line) obtained letting $A_{\rm amp}$
to be a free parameter; the fit is obtained for $A_{\rm amp}=1.86$.
It is indeed similar with the best-fit non-Gaussian curve, but there
is no physical justification for $A_{\rm amp}\neq 1$.
\label{fig:qso}}
\end{center}
\end{figure*}

In spite of the high efficiency of the selection algorithm adopted
to define the SDSS DR6 QSO catalog, some contamination from
UV-excess stars is unavoidable. Following ref. \cite{myersetal06},
we model the observed ACF, $\hat{c}^{\rm tt}$, as the sum of
contributions from QSOs, $\hat{c}^{\rm qq}$, and from stars,
$\hat{c}^{\rm ss}$, plus an offset, $\epsilon$, arising from
cross-terms:
\begin{equation}
\hat{c}^{\rm tt}(\theta)= a^2\hat{c}^{\rm
qq}(\theta)+(1-a)^2\hat{c}^{\rm ss}(\theta) +
\epsilon(\theta)~,\label{crossterm}
\end{equation}
where $a$ is the efficiency of the QSO catalog, i.e. the fraction of
true QSOs. As shown by ref. \cite{myersetal06}, the KDE
classification technique is efficient enough for the offset
$\epsilon$ to be safely neglected. The contributions from stars and
QSOs can be disentangled exploiting their different dependencies on
$\theta$. In fact refs. \cite{myersetal06} and \cite{xiaisw} showed
that $\hat{c}^{\rm ss}(\theta)$ keeps almost flat up to large
angular scales, and is therefore expected to dominate the signal on
scales of several degrees. We have checked that the contribution
from contaminating stars to the QSO-CMB CCF can safely be neglected.

As in the NVSS case we used the HEALPix software to pixelize the QSO
map at the $N_{\rm side}=64$ resolution and the jackknife
re-sampling method to estimate the covariance matrix among data
points. Our estimates of the ACF for the SDSS DR6 QSO catalogue and
of the QSO-CMB CCF, shown in figure \ref{fig:qso}, are consistent
with previous analyses \cite{gianna08,xiaisw}. Note that we do not
split the SDSS QSO in different redshift bins using the objects'
photometric redshift estimates, but we consider the projected ACF
(and CCF) of the full sample.

\section{Method \&  Data Analysis}\label{method}

We perform a global fitting of cosmological parameters, including
$f_{\rm NL}$, for the data of Sec. \ref{data} including also the
datasets described below, using the {\tt CosmoMC}
\footnote{Available at http://cosmologist.info/cosmomc/.} package
\cite{cosmomc}, a Markov Chain Monte Carlo (MCMC) code, modified to
calculate the theoretical ACF and CCF. We assume purely adiabatic
initial conditions and a flat Universe, with no tensor contribution
to primordial fluctuations. The following six cosmological
parameters are allowed to vary with top-hat priors: the dark matter
energy density parameter $\Omega_{\rm c} h^2 \in [0.01,0.99]$, the
baryon energy density parameter $\Omega_{\rm b} h^2 \in
[0.005,0.1]$, the primordial spectral index $n_{\rm s} \in
[0.5,1.5]$, the primordial amplitude $\log[10^{10} A_{\rm s}] \in
[2.7,4.0]$, the ratio (multiplied by 100) of the sound horizon at
decoupling to the angular diameter distance to the last scattering
surface $\Theta_{\rm s} \in [0.5,10]$, and the optical depth to
reionization $\tau \in [0.01,0.8]$. The pivot scale is set at
$k_{\rm s0}=0.05\,$Mpc$^{-1}$ and do not consider massive neutrinos
and dynamical dark energy.

Besides these six basic cosmological parameters, we have three more
parameters related to the ACF and CCF data: the non-Gaussianity
parameter $f_{\rm NL}$, the minimal halo mass $M_{\rm min}$, two
offset parameters $c$, one for the ACF and one for the CCF, allowing
for the effect of the integral constraint. The ranges for $f_{\rm
NL}$ and $M_{\rm min}$ are the same as in ref. \cite{xiaACF}. Those
for the offsets were chosen to be large enough to account for the
large-scale signals. Specifically we have, for the NVSS: $c_{\rm
ACF} \in [10^{-8},10^{-4}]$ (see \S\,\ref{ACFdata}), $c_{\rm CCF}
\in [-0.5,0.5]$; for the QSOs: $c_{\rm ACF} \in [-2\times
10^{-3},2\times 10^{-3}]$, $c_{\rm CCF} \in [-0.5,0.5]$. All the
best fit values of these parameters are well within these ranges and
far from the boundary values (see \S\,\ref{result}).

In the QSO case we have, in addition, the efficiency of QSO
classification $a$. Several authors also treated as an additional
free parameter the ISW amplitude $A_{\rm amp}$ defined by
$\bar{c}^{\rm gT}(\theta)=A_{\rm amp}c^{\rm gT}(\theta)$, where
$\bar{c}^{\rm gT}$ and $c^{\rm gT}$ are the observed and theoretical
CCF. We note that values of $A_{\rm amp}\neq 1$ have no physical
meaning, and are indicative of some unrecognized problem in the
interpretation of the data. Therefore we fix $A_{\rm amp}= 1$ in all
our calculations, except to estimate the significance of the ISW
signal, in order to compare with previous works in which this
parameter was used.

The model ACF $c^{\rm gg}(\theta)$ and CCF $c^{\rm gT}(\theta)$ are
compared with the observed values $\hat{c}^{\rm gg}(\theta)$ and CCF
$\hat{c}^{\rm gT}(\theta)$, respectively, through the Gaussian
likelihood function:
\begin{eqnarray}
\mathcal{L}_{\rm ACF}&=&(2\pi)^{-N/2}[{\rm
det}C_{ij}]^{-1/2}\times\exp\left[-\sum_{i,j}\frac{C^{-1}_{ij}(\hat{c}^{\rm
gg}_i-{c}^{\rm gg}_i)(\hat{c}^{\rm gg}_j-{c}^{\rm gg}_j)}{2}\right]~,\\
\mathcal{L}_{\rm CCF}&=&(2\pi)^{-N/2}[{\rm det}C'_{ij}]^{-1/2}\times
\exp\left[-\sum_{i,j}\frac{C'^{-1}_{ij}(\hat{c}^{\rm gT}_i-{c}^{\rm
gT}_i)(\hat{c}^{\rm gT}_j-{c}^{\rm gT}_j)}{2}\right]~,
\end{eqnarray}
where $C_{ij}$ and $C'_{ij}$ are the observed ACF and CCF
covariance matrices.


The following cosmological data are also included in the fit: ${\rm
i})$ power spectra of CMB temperature and polarization anisotropies;
${\rm ii})$ baryonic acoustic oscillations (BAOs) in the galaxy
power spectra; ${\rm iii})$ SNIa distance moduli.

To deal with the 7 years WMAP (WMAP7) CMB temperature and
polarization power spectra we use the routines for computing the
likelihood supplied by the WMAP team \cite{WMAP7}. The WMAP7 data
are exploited only to improve the constraints on the six basic
cosmological parameters, not to constrain $f_{\rm NL}$.

The BAOs \cite{BAO} can, in principle, measure not only the angular
diameter distance, $D_A(z)$, but also the expansion rate of the
Universe, $H(z)$. However, the limited accuracy of current data only
allows us to determine the ratio between the distance scale defined
by ref. \cite{Eisenstein:2005su}:
\begin{equation}
D_v(z)\equiv\left[(1+z)^2D_A^2(z)\frac{cz}{H(z)}\right]^{1/3},
\end{equation}
and the comoving sound horizon at the baryon-drag epoch, $r_s(z_d)$
(see ref. \cite{Eisenstein:1997ik}). Accurate determinations of the
distance ratio $r_s(z_d)/D_v(z)$, $r_s(z_d)$ have been obtained by
ref. \cite{BAO}:
\begin{eqnarray}
r_s(z_d)/D_v(z=0.20)&=&0.1905\pm0.0061,\nonumber\\
r_s(z_d)/D_v(z=0.35)&=&0.1097\pm0.0036.
\end{eqnarray}
We adopt these values as Gaussian priors.

The SNIa data yield the luminosity distance as a function of
redshift which provides strong constraints on the dark energy
evolution. We use the Union compilation data (307 samples) from the
Supernova Cosmology project \cite{Kowalski:2008ez}, which include
the samples of SNIa from the (Supernovae Legacy Survey) SNLS and
from the ESSENCE survey, and span the redshift range
$0\lesssim{z}\lesssim1.55$. In the calculation of the likelihood
from SNIa we marginalize over the nuisance parameter as done in
refs. \cite{SNMethod1,SNMethod2}.

Furthermore, we add a prior on the Hubble constant, $H_0=74.2 \pm
3.6$ km/s/Mpc given by ref. \cite{HST}. Finally, in the analyses of
NVSS sources and SDSS QSOs we set the minimal halo mass at $M_{\rm
min} > 10^{12} h^{-1}M_{\odot}$ consistent with the results of ref.
\cite{Croom}.

\begin{figure}
\begin{center}
\includegraphics[scale=0.43]{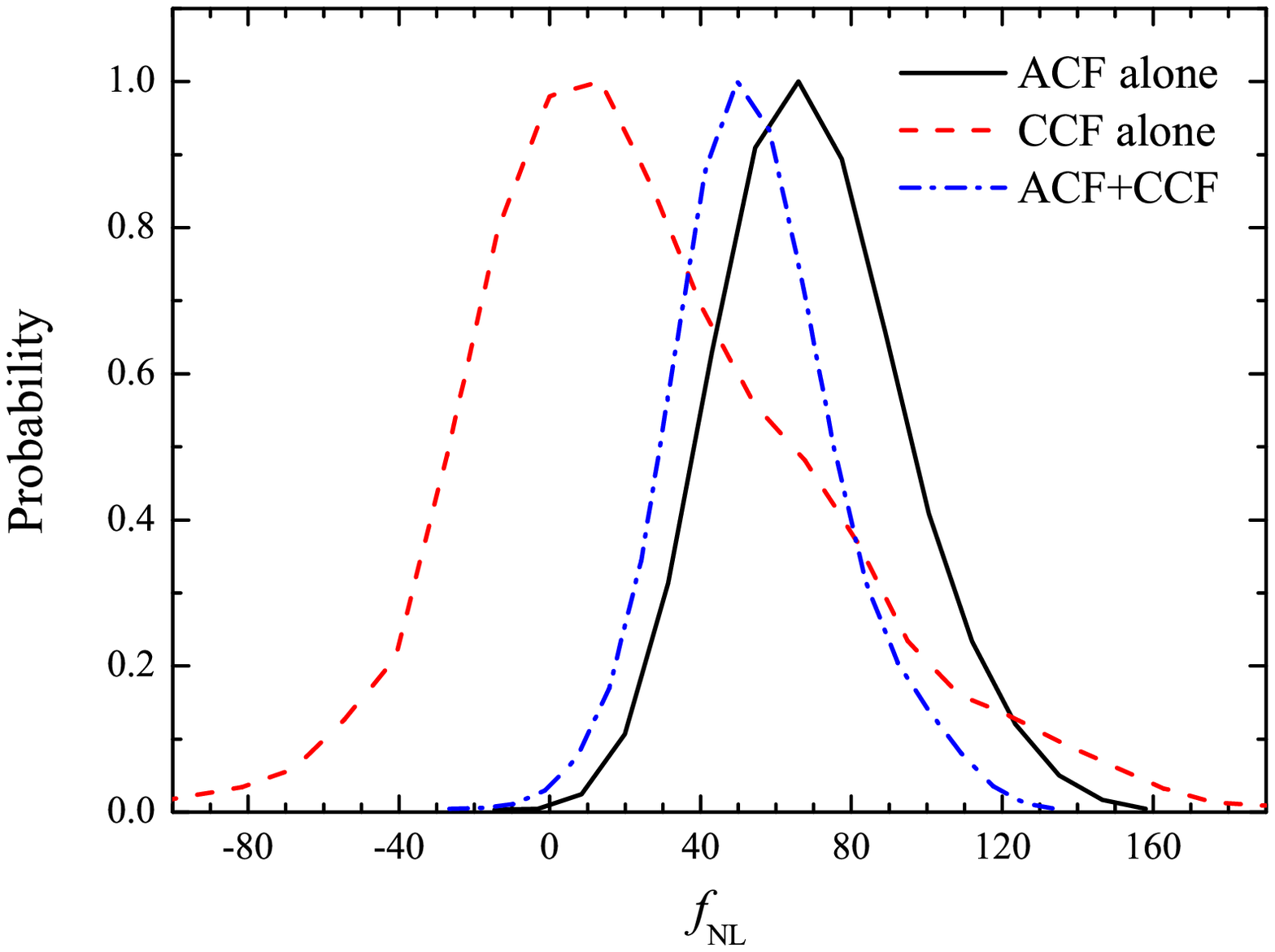}
\caption{Marginalized one-dimensional posterior distribution of the
non-Gaussianity parameter $f_{\rm NL}$ from different NVSS data
combinations: ACF alone (black continuous curve), CCF alone (red
dashed curve) and ACF+CCF (blue dot-dashed curve).
\label{fig:nvss1d}}
\end{center}
\end{figure}

\section{Numerical Results}\label{result}

\subsection{NVSS Radio Sources}\label{resNVSS}

To check the significance of the detection of the late-time ISW
effect from the NVSS-CMB CCF, following ref. \cite{xiaisw}, we
compute the best fit value  of the amplitude $A_{\rm amp}$ in the
Gaussian case ($f_{\rm NL}=0$) and the jackknife error. We obtain:
\begin{equation}
A_{\rm amp}=1.14 \pm 0.50~~~(1\,\sigma~{\rm error})~,
\end{equation}
so that the significance is $\simeq A_{\rm amp}/\sigma_{\rm A}$,
i.e. $\simeq 2.3\,\sigma$, broadly consistent with earlier analyses
\cite{gianna08,Raccanelli,Massardi10} and very well compatible with
its physical value (unlike ref. \cite{hoetal08}).

The constraints on $f_{\rm NL}$, obtained setting $A_{\rm
amp}\equiv1$ and marginalizing over all the other parameters, are
summarized in Table~\ref{tab:I}. The NVSS CCF turns out to be only
weakly sensitive to the minimum halo mass. Therefore we fix it to
the best fit value from NVSS ACF data alone, $M_{\rm
min}=10^{12.47}h^{-1}M_{\odot}$. While the NVSS ACF yields a
positive {$f_{\rm NL}$} at more than $2\,\sigma$ confidence level,
$f_{\rm NL}=58\pm 28$ ($1\,\sigma$), as previously found by ref.
\cite{xiaACF}, the NVSS-CMB CCF provides much weaker constraints:
$f_{\rm NL}=29\pm 48$ ($1\,\sigma$). The best fit value for the
offset is $c\simeq-0.05$ for $S_{1.4\rm GHz}\ge20$mJy. To check the
effect of residual systematic fluctuations in the NVSS source
density, we have repeated the calculation for $S_{1.4\rm GHz}\ge
10\,$mJy. We find that the amplitude of the CCF decreases,
consistent with the cross-correlation being somewhat blurred (a
similar effect can be gleaned also from figure 11 of ref.
\cite{CHM10}), and the constraints on {$f_{\rm NL}$} become $f_{\rm
NL}=6\pm 37$ ($1\,\sigma$).




Other systematic effects may affect the CCF estimates (for a
parallel analysis on the ACF see ref. \cite{xiaACF}). A first
possibility is that the CMB map is contaminated by residual
foreground emissions that increase the noise and thus swamp the
signal we are looking for. The contamination from point sources
below the detection limit is increasingly diluted for larger and
larger pixel sizes. However we find that the CCF estimate does not
change in any appreciable way when we use larger pixel sizes
($N_{\rm side} = 32$ and $N_{\rm side} = 16$). Note that increasing
the pixel size we also decrease the relative contribution of CMB
fluctuations generated at the recombination, which have much higher
amplitude than those due to the ISW effect.

\begin{figure}
\begin{center}
\includegraphics[scale=0.45]{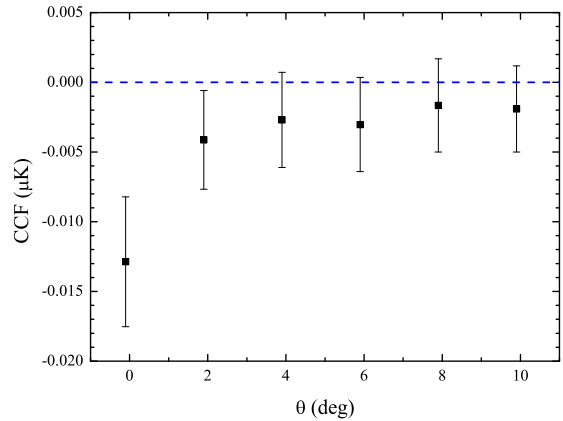}
\caption{Cross-correlation between the ILC CMB and a Galactic
emission template comprising synchrotron, free-free and dust
contributions (black squares). The Galactic template is our best
guess of the residual Galactic contamination in the ILC obtained as
described in ref. \cite{Stivoli}. Error bars are
jackknife-estimated. \label{fig:foreground}}
\end{center}
\end{figure}

The impact of residual contamination from diffuse Galactic emissions
is assessed by cross-correlating the CMB map with the standard dust,
free-free and synchrotron templates, used also by the WMAP team
(e.g., ref. \cite{Gold}), smoothed to 1 deg resolution. No
indications of significant cross-correlations are found. However,
residual foreground contamination will be due a certain mixture of
the different Galactic emissions, thus it is not expected to
strongly correlate with single foreground templates. We thus build a
typical foreground residual map to be correlated with the CMB map,
as described in the following.

Ref. \cite{Stivoli} shows how to predict the residual contamination
after the application of a linear component separation technique.
The method requires to have a model of the data and to know the
weights of the linear mixture for component separation.

Our data model is composed as follows: for the frequency
dependencies of the components is based on model M2 of ref.
\cite{Bonaldi}, obtained from an analysis of the WMAP 3 years data;
for the spatial morphology is based on the previously mentioned
foreground templates. The component-separation linear-mixture
weights are the ILC weights adopted to build the CMB map, available
from the LAMBDA site.


The foreground residual map we obtain is found to have a
statistically significant negative CCF with the ILC map (figure
\ref{fig:foreground}), indicating that the separation of the
Galactic emission from the CMB map is not perfect. The amplitude of
this spurious cross-correlation  is however too small to affect
significantly the NVSS-CMB CCF.

If we combine with the NVSS ACF and CCF data we obtain the following
constraint on $f_{\rm NL}$:
\begin{equation}
f_{\rm NL}=53\pm25~~~(1\,\sigma~{\rm error}),
\end{equation}
or $10<f_{\rm NL}<106$ at $95\%$ confidence level. This result is
compatible with previous estimates
\cite{Yadav:2007yy,Pietrobon:2008ve,Slosar,Curto:2009pv,Smidt:2009ir,JV,Smith:2009jr,Rud},
and with the WMAP7 limits \cite{WMAP7}.

\begin{figure}[t]
\begin{center}
\includegraphics[scale=0.35]{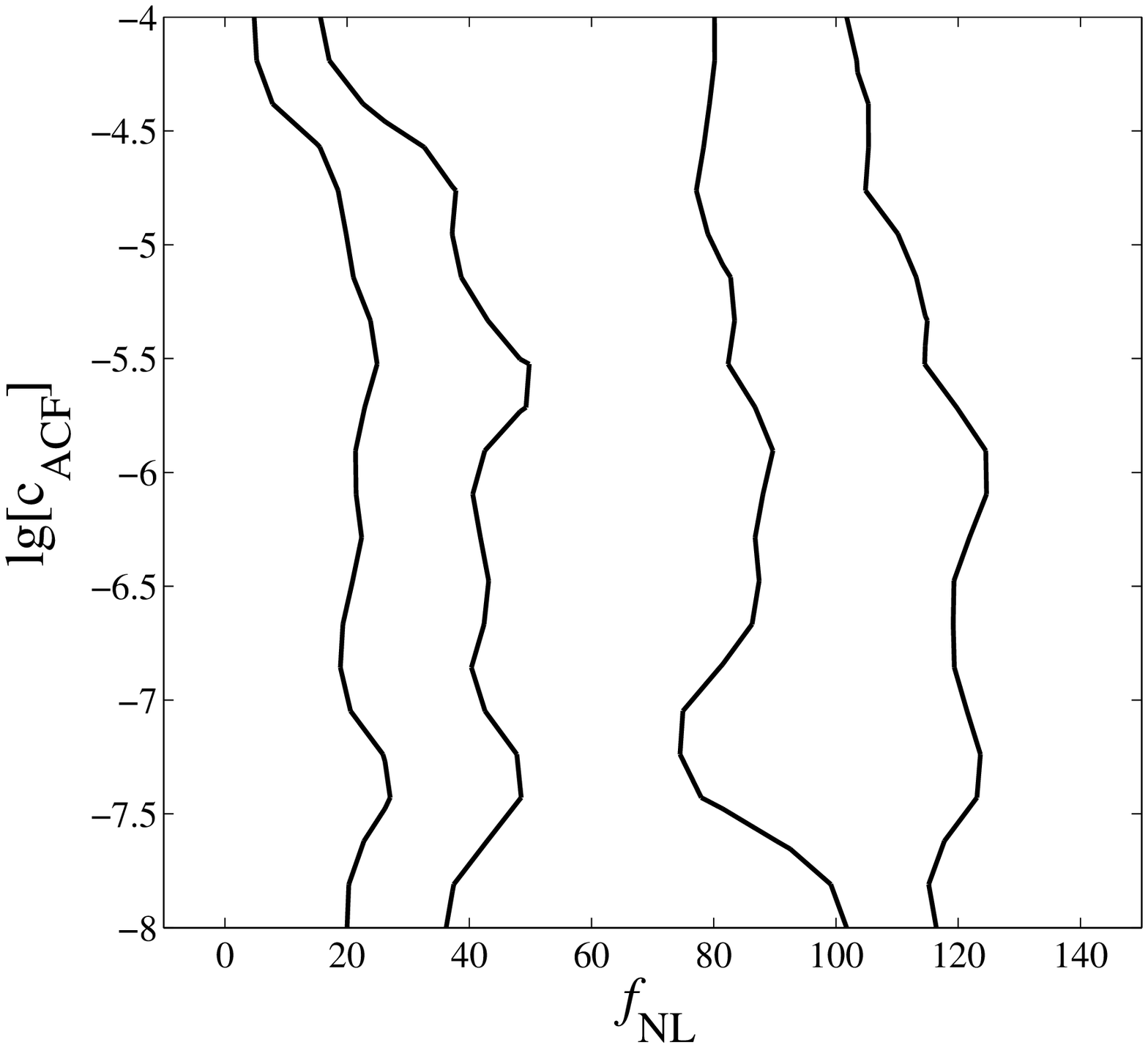}
\includegraphics[scale=0.35]{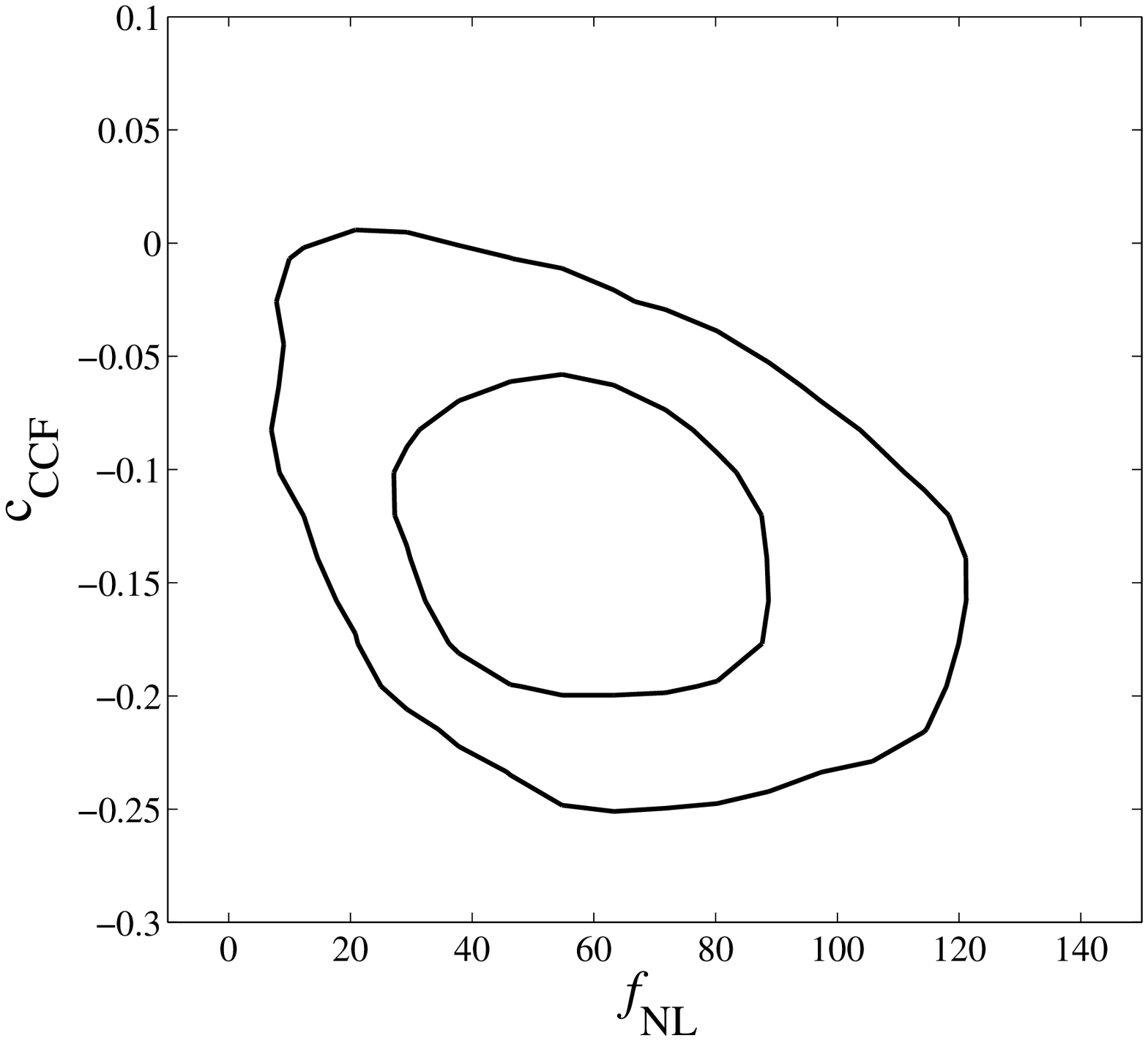}
\caption{Two-dimensional marginalized 1 \& $2\,\sigma$ contours in the ($f_{\rm
NL},\log_{10}[c_{\rm ACF}]$) and ($f_{\rm NL},c_{\rm CCF}$) from the
NVSS ACF and CCF data.} \label{fig:nvss2d}
\end{center}
\end{figure}

In figure \ref{fig:nvss2d} we plot the two-dimensional constraints
on ($f_{\rm NL},c_{\rm ACF}$) and ($f_{\rm NL},c_{\rm CCF}$) from
the NVSS ACF and CCF data. As noted above, a positive offset means
that the true ACF or CCF is lower than the estimated one, thus
weakening or spoiling indications for non-Gaussianity. On the
contrary, a negative offset points to larger values of $f_{\rm NL}$.
As discussed in section~\ref{ACFdata}, the integral constraint
should have a negligible effect on the NVSS ACF but an offset of up
to $10^{-4}$ may be induced by large scale density gradients of
instrumental origin. This value has been adopted as the upper
boundary for the allowed range for $c_{\rm ACF}$.

\begin{table}
\caption{$1,\,2\,\sigma$ constraints on the primordial
non-Gaussianity from different data combinations. We report the mean
values and the Bayesian central credible interval, marginalized over
all other parameters.} \label{tab:I}
\begin{center}
\begin{tabular}{c|c|c}
\hline \hline

Datasets&\multicolumn{2}{c}{Non-Gaussianity $f_{\rm NL}$}\\

\hline

WMAP7+BAO+SN&~~~~$1\,\sigma$ C. L.~~~~&~~~~$2\,\sigma$ C. L.~~~~\\

\hline

\multicolumn{3}{c}{NVSS Radio Sources}\\

\hline
+ACF&$58\pm28$&$[16,114]$\\
+CCF&$29\pm48$&$[-50,145]$\\
+ACF+CCF&$53\pm25$&$[10, 106]$\\


\hline\hline

\multicolumn{3}{c}{SDSS DR6 QSOs}\\

\hline
+ACF&$42\pm28$&$[-19,93]$\\
+CCF&$60\pm42$&$[-20,145]$\\
+ACF+CCF&$58\pm24$&$[12,94]$\\

\hline \hline
\end{tabular}
\end{center}
\end{table}

\begin{figure}[t]
\begin{center}
\includegraphics[scale=0.45]{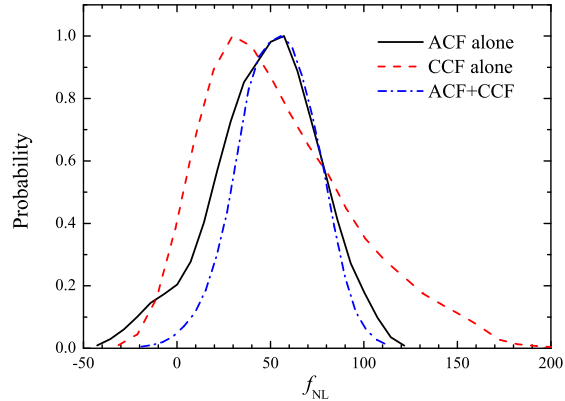}
\caption{Marginalized one-dimensional posteriors of the
non-Gaussianity parameter $f_{\rm NL}$ from different SDSS QSO data
combinations: ACF alone (black continuous curve), CCF alone (red
dashed curve) and ACF+CCF (blue dot-dashed curve). See text for more
details.} \label{fig:qso1d}
\end{center}
\end{figure}

\subsection{SDSS DR6 QSOs}

As in Section \ref{resNVSS}, to assess the significance of the
detection of a cross-correlation between the distribution of SDSS
DR6 QSOs and the CMB map we compute the best fit value of $A_{\rm
amp}$ and its error keeping all the cosmological parameters fixed at
the WMAP7 values. We find:
\begin{equation}
A_{\rm amp}=1.86 \pm 0.80~~~(1\,\sigma~{\rm error})~,
\end{equation}
implying a $\sim 2.3\,\sigma$ detection of the late-time ISW
effect, consistent with previous analyses.

The constraints on {$f_{\rm NL}$} from the ACF, the CCF and their
combination are given in Table~\ref{tab:I} and illustrated in
figure~\ref{fig:qso1d}. As for NVSS sources, these constraints are
obtained setting $A_{\rm amp}=1$ and marginalizing over all the
other parameters. Allowing for the integral constraint may be very
important for this data set, as the surveyed area is only $\simeq
1/4$ of the NVSS area and its effective angular size 
is not much larger than the scale 
over which positive correlations are induced by the {$f_{\rm NL}$}
value indicated by the NVSS ACF data. As for the NVSS case we added
constants $c$, as free parameters, to the ACF and to the CCF, and
marginalized over them. In the ACF case, the effect of this constant
is negligible. Its best fit value is $c_{\rm ACF}= -2\cdot 10^{-4}$,
and we find $f_{\rm NL}= 42\pm 28$ ($1\,\sigma$); if we set $c_{\rm
ACF}=0$ we get $f_{\rm NL}= 32\pm 19$ ($1\,\sigma$). As expected,
the impact of the constant offset is much more relevant for the CCF.
We find $c_{\rm CCF}=-0.15$ and $f_{\rm NL}= 60\pm 42$
($1\,\sigma$); setting $c_{\rm CCF}=0$ we get $f_{\rm NL}= 0\pm 10$
($1\,\sigma$). Clearly marginalizing over the $c_{\rm CCF}$
parameter has a big effect on the recovered $f_{\rm NL}$ value. The
larger the sky coverage of the survey the better this constant is
determined and the less is its impact on $f_{\rm NL}$.

Combining the ACF and CCF data we find $f_{\rm NL}= 58\pm 24$
($1\,\sigma$), $c_{\rm ACF}= -3\cdot 10^{-4}$, $c_{\rm CCF}=-0.13$.
The value of {$f_{\rm NL}$} is nicely consistent with that obtained
from the analysis of NVSS data. The best fit value of the selection
efficiency of the QSO catalogue, kept as a free parameter, is
$a=97.1\%\pm 1\%$, consistent with the estimate by ref.
\cite{richards09} who claim an efficiency of over 97\% for the UVX
sub-sample considered here.

\begin{figure}[t]
\begin{center}
\includegraphics[scale=0.35]{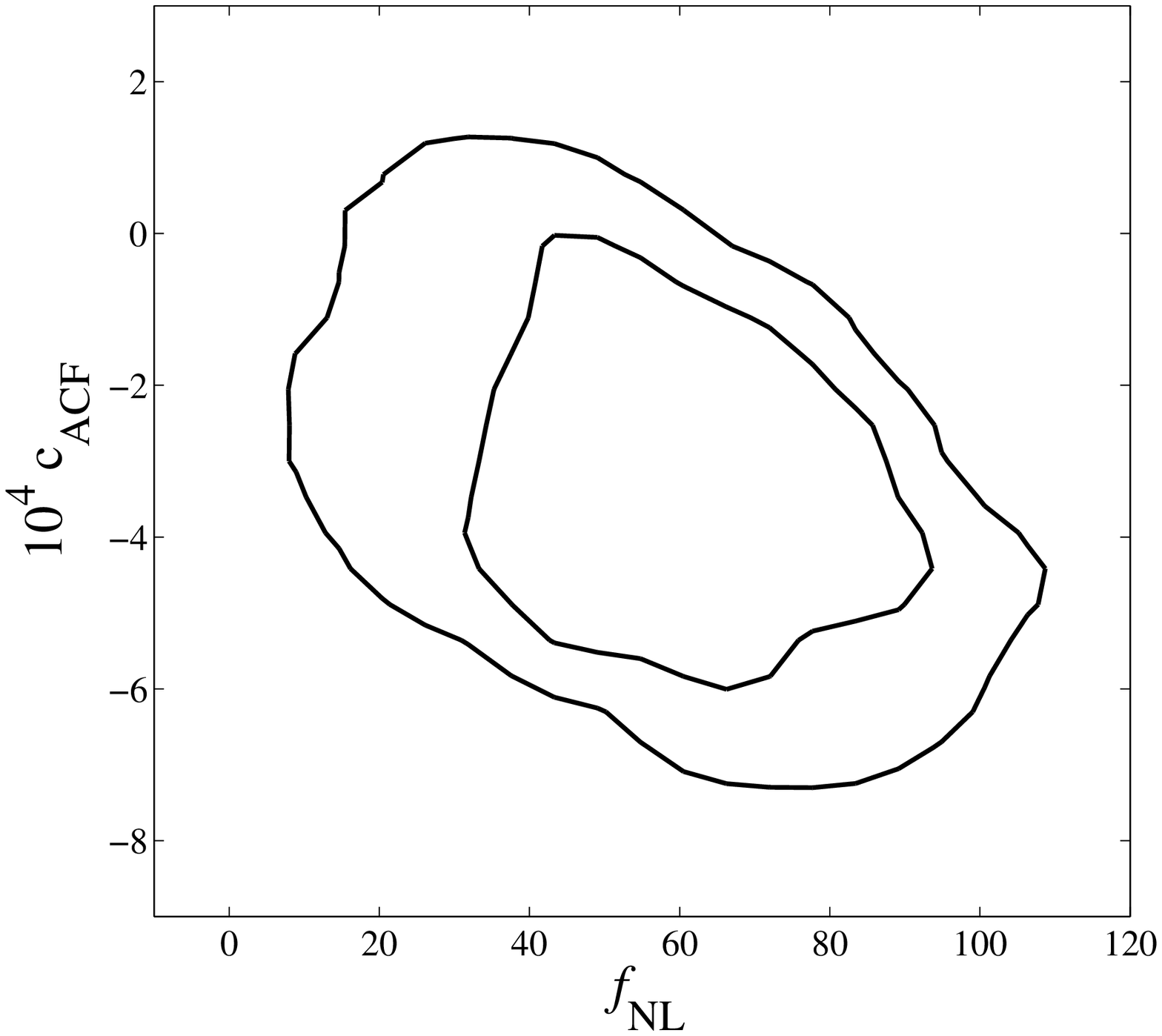}
\includegraphics[scale=0.35]{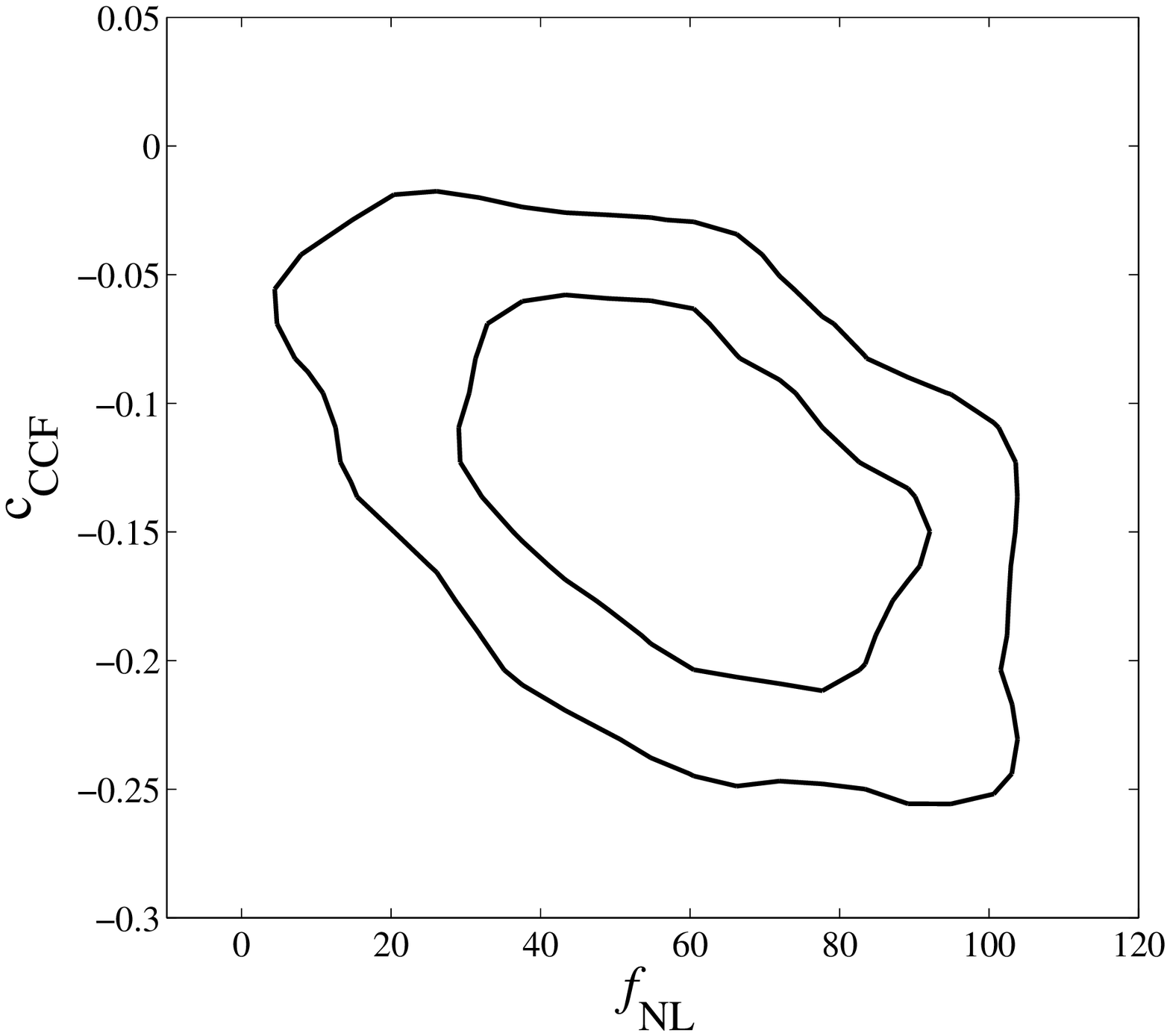}
\caption{Two-dimensional marginalized  1 \& $2\,\sigma$ contours in
the ($f_{\rm NL},10^4c_{\rm ACF}$) and ($f_{\rm NL},c_{\rm CCF}$)
from the SDSS QSO ACF and CCF data.} \label{fig:qso2d}
\end{center}
\end{figure}

Finally, in figure \ref{fig:qso2d} we plot the two-dimensional
constraints on ($f_{\rm NL},10^4c_{\rm ACF}$) and ($f_{\rm
NL},c_{\rm CCF}$) from the SDSS QSO ACF and CCF data. Both $c_{\rm
ACF}$ and $c_{\rm CCF}$ are anti-correlated with $f_{\rm NL}$. A
negative $c_{\rm ACF}$ and $c_{\rm CCF}$ enhances the estimate of
$f_{\rm NL}$.

\section{Discussion \& Conclusions}\label{summary}

All the datasets we considered here yield broadly consistent
results. The NVSS ACF drives the positive signal $16<f_{\rm NL}<114$
at $95\%$ confidence level. The SDSS QSO ACF and CCF individually
yield results consistent with $f_{\rm NL}=0$.  This effect arises
because of degeneracies in a high-dimensional parameter space.  In
particular, there  is a degeneracy  in the $M_{\rm min}-f_{\rm NL}$
plane, with large values of  $M_{\rm min}$ corresponding to low
$f_{\rm NL}$ values, which, when one marginalizes over $M_{\rm min}$
lowers the mean $f_{\rm NL}$ value. The combination ACF+CCF however
seems to break this degeneracy (ACF depends on $b^2$ while CCF
depends on $b$) cutting out the large $M_{\rm min}$ tail and thus
increasing the recovered $f_{\rm NL}$ value. In addition the
relatively large and negative $c_{\rm CCF}$ skews the maximum
posterior $f_{\rm NL}$ value for the CCF to larger positive values.
Before drawing our conclusions we compare our findings with previous
work in the literature where very similar data-sets were used.

\subsection{Comparison with Previous Works}

The work presented here is closely related to refs.
\cite{Slosar,afshorditolley} and to the NVSS-CMB cross power
spectrum as obtained by ref. \cite{hoetal08}. The most immediate
difference in the analysis is that we work with the ACF and CCF
while previous works has used the auto and cross power spectra.
Correlation functions and power spectra are spherical harmonics
transform pairs and should thus be fully equivalent. The two
approaches however are effected by systematic effects in slightly
different ways: whenever possible we try to directly compare the two
approaches. In addition, compared with previous analyses, we
marginalize over a suite of possible systematic effects, modeled by
the constant offsets $c_{\rm ACF}$ and $c_{\rm CCF}$ and over a
possible stellar contamination and QSO efficiency parameter.

Ref. \cite{afshorditolley} used the NVSS-WMAP 5 year map cross power
spectrum as provided by ref. \cite{hoetal08} to obtain $f_{\rm
NL}=236\pm 127$ ($1\,\sigma$). The main differences with our NVSS
CCF analysis are: {\it i}) ref. \cite{afshorditolley} used a
constant bias, here we use eq. (\ref{eq:bST}), we include the
non-Gaussianity correction in the mass function, thus have a free
parameter given by the minimum host halo mass for the sources;  {\it
ii}) refs. \cite{afshorditolley,hoetal08} used WMAP 5 years map,
while here we use up-to-date cleaned CMB maps;  {\it iii}) ref.
\cite{hoetal08} use NVSS sources much dimmer ($\ge2.5$ mJy) than our
sample ($\ge20$ mJy). They have a higher source density but we have
seen that density gradients are most important for dim sources.
Finally, we do not attempt to model the redshift distribution of
NVSS sources but use directly the one that has been measured; this
results in our errors on $f_{\rm NL}$ to be smaller, but the two
results are fully consistent.

Ref. \cite{Slosar} obtained non-Gaussianity constraints using both
the NVSS and SDSS QSO data. For their NVSS analysis they also used
it only in cross correlation with the CMB (they do not use the NVSS
sources power spectrum) and, like ref. \cite{afshorditolley}, used
the cross-correlation data and errors of ref. \cite{hoetal08}. In
their analysis the combinations: bias times source redshift
distribution, $b\cdot dn/dz$, and the bias as function of redshift,
$b(z)$, as well as $f_{\rm NL}$ are determined from the correlation
function data themselves, here instead we use the actual redshift
distribution and use the minimum halo host mass as a parameter,
rather than $b(z=0)$. In their reported ISW results several data
sets are combined: the SDSS QSOs, the NVSS, 2MASS etc. although
they report their results to be dominated by NVSS. Our NVSS CCF
constraints on $f_{\rm NL}$ are consistent with theirs although with
smaller error-bars due to the fact that we do not need to estimate
the source redshift distribution from the correlations themselves.

For their SDSS QSO analysis ref. \cite{Slosar} considered both the
auto and cross correlation signals. Ref. \cite{Slosar} used an
extension of the DR3 QSO sample that include many of the sources
that subsequently were released with the DR6 sample: we use the more
complete and better calibrated final official SDSS DR6 QSO catalog
release \cite{richards09}. Ref. \cite{Slosar} found that their QSO
sample at $z<1.45$ seem to suffer from contamination which they
described to systematic calibration errors, and discarded this
sample from their analysis. We have checked for this effect and find
no evidence in the sample we use that the $z<1.45$ QSOs have
different contamination than the $z>1.45$ or that the two samples
give different estimates of $f_{\rm
NL}$. Our sample is more than double the size than theirs.

In our analysis we do not split the QSO in redshift bins according
to each source photometric redshift estimate as ref. \cite{Slosar}
do, but we consider the projected correlation of the full sample.
Ref. \cite{Slosar} had therefore to estimate the source redshift
distribution from the (cross)correlation signals themselves for each
of their sub-samples; we avoid this by using the
independently-estimated full sample source redshift distribution. We
also leave the stellar contamination as a free parameter to
marginalize over, while ref. \cite{Slosar} kept it fixed at the
fiducial value.

The two analyses also use different bias models: ref. \cite{Slosar}
used a functional form $b(z)\sim (1+z)^5$ or $b(z)\sim 1/D(z)$ with
different constant offset depending on assumptions  about ``recent
merging'' activity. We use the extended Press-Schecter bias
formulation of eq. (\ref{eq:bST}). We have however verified that if
we use their bias model, fix the QSO efficiency (i.e. fraction of
stellar contamination in the catalog) to the fiducial value and do
not marginalize over the constant offset we recover even central
$f_{\rm NL}$ values fully consistent with theirs ($f_{\rm NL}=6 \pm
18$ ($1\,\sigma$)). We therefore conclude that the largest effect
driving the different central $f_{\rm NL}$ values we recover is the
treatment of systematic effects -stellar contamination and offset.

We should stress here that for the data in common between the
different analyses, the results we find are fully consistent (i.e.,
at the $1\,\sigma$ confidence level) with both refs.
\cite{afshorditolley} and \cite{Slosar} results. Our positive
$f_{\rm NL}$ signal is driven by the NVSS auto-correlation function
and by the combination of SDSS DR6 QSO ACF  and CCF where the
combination breaks the degeneracy with $M_{\rm min}$ pushing he
marginalized $f_{\rm NL}$ values slightly up.

Better knowledge of the bias of the sample and of systematic effects
such as stellar contamination, calibration and  selection effects
are  clearly needed to make  further  progress on this front; this
is especially relevant to future data-sets where the large volume
surveyed will further reduce the statistical errors.

\subsection{Conclusions}
We have investigated the constraints on the parameter $f_{\rm NL}$,
characterizing primordial non-Gaussianity of the local type, coming
from the currently available data sets best suited for this purpose,
namely the observationally determined auto-correlation functions
(ACFs) of NVSS radio sources and the SDSS-DR6 QSOs, and the
cross-correlation functions (CCFs) of the surface density
distributions of these sources with the WMAP 7 years CMB map. The
mean redshift of NVSS sources is $\bar{z} \simeq 1.23$, and that of
SDSS-DR6 QSOs is $\bar{z}\simeq1.49$.

As shown by ref. \cite{xiaACF}, a positive {$f_{\rm NL}$} value
($f_{\rm NL}=58\pm28$ ($1\,\sigma$), after allowing for the possible
effect of the integral constraint), consistent with the constraints
set by previous analyses, and especially by WMAP data \cite{WMAP7}
($f_{\rm NL}=32\pm21$ ($1\,\sigma$)), can explain the positive ACF
signal detected on angular scales $> 4^\circ$ (see refs.
\cite{BlakeWall,Overzier}), where the ACF is expected to be $\le 0$
for Gaussian primordial perturbations. The QSO ACF yields
constraints ($f_{\rm NL}=42\pm28$ ($1\,\sigma$)) very similar to
those inferred from WMAP 7 year data.

A signature of $f_{\rm NL}> 0$ should also be present in the CCF
between tracers of large scale structure and the map of CMB
anisotropies. Most of the signal, however, comes from low multipoles
($\ell <10$) corresponding to very large angular scales where the
signal may be blurred by systematic effects. In the case of NVSS
sources we have indeed found that the amplitude of the CCF with the
CMB increases somewhat as we increase the adopted flux limit from
$S_{1.4\rm GHz}=10\,$mJy, which is found to be adequate for the ACF
analysis but is still affected by small spurious large scale surface
density gradients, to $S_{1.4\rm GHz}=20\,$mJy where such gradients
become negligible. Other possible systematic effects, such as
contamination of the CMB map by point sources below the detection
limit or by residual Galactic emissions have been shown to be of
minor importance. The NVSS-CMB CCF however turned out to provide
rather loose constraints on {$f_{\rm NL}$} ($f_{\rm NL}=29\pm 48$
($1\,\sigma$)).

The limited area covered by the SDSS-DR6 QSO catalog implies that
the CCF is liable to a quite substantial effect of the integral
constraint, at variance with the NVSS case, where the area is 4
times larger (the ACFs of both NVSS sources and SDSS QSOs are found
to be insensitive to this effect). We have allowed for the integral
constraint by adding to the estimated CCF a constant offset, treated
as a free parameter, and marginalizing over it. In this way we
obtained $f_{\rm NL}=60\pm 42$ ($1\,\sigma$).

Finally, we obtain the constraints on $f_{\rm NL}=53\pm25$
($1\,\sigma$) and $f_{\rm NL}=58\pm24$ ($1\,\sigma$) from NVSS data
and QSO data, respectively. The tantalizing hints of a positive
{$f_{\rm NL}$} reported by ref. \cite{xiaACF} have survived the
present more extensive analysis, involving other data sets. In order
to reconcile the data with $f_{\rm NL}=0$, we need to show that
either the NVSS catalog is affected by many-times larger systematic
effects (spurious large-scale correlations  introduced by density
gradients) than known and quantified so far, or that the redshift
evolution of the bias  of these sources  is radically different from
that of their radio-quiet, optical counterpart.

Alternatively, one may question the accuracy of jackknife
error-estimation especially when there are many non-zero
off-diagonal elements in the covariance matrix. From ref.
\cite{scranton03}, we infer that jackknife can underestimate
parameter errors by up to 30\%. If the error bars were to be
increased, {$f_{\rm NL}$} would become more compatible with zero.
Much tighter and robust constraints on {$f_{\rm NL}$} should be
provided by the forthcoming surveys with ASKAP \cite{ASKAP} and
MeerKAT \cite{MeerKAT}, the SKA pathfinders, and by much larger area
QSO surveys.

\section*{Acknowledgements}

We are indebted to Jasper Wall and Chris Blake for clarifications on
their analysis of the NVSS ACF and for useful comments, to Gordon T.
Richards for clarifications about the SDSS DR6 QSO catalog, to
Marcella Massardi for help with constructing the CMB mask described
in \S\,\ref{ACFdata}, and to an anonymous referee for a quick and
helpful report. We acknowledge the use of the Legacy Archive for
Microwave Background Data Analysis and the HEALPix package.
Numerical analysis has been performed at the University of Cambridge
High Performance Computing Service (http://www.hpc.cam.ac.uk/). This
research has been partially supported by the ASI contract No.
I/016/07/0 COFIS, the ASI/INAF agreement I/072/09/0 for the Planck
LFI Activity of Phase E2 and a PRIN MIUR, MICCIN grant AYA2008-0353,
FP7-IDEAS-Phys. LSS 240117, FP7-PEOPLE-2007-4-3-IRGn202182. Funding
for the SDSS and SDSS-II has been provided by the Alfred P. Sloan
Foundation, the Participating Institutions, the National Science
Foundation, the U.S. Department of Energy, the National Aeronautics
and Space Administration, the Japanese Monbukagakusho, the Max
Planck Society, and the Higher Education Funding Council for
England. The SDSS Web Site is http://www.sdss.org/. The SDSS is
managed by the Astrophysical Research Consortium for the
Participating Institutions. The Participating Institutions are the
American Museum of Natural History, Astrophysical Institute Potsdam,
University of Basel, University of Cambridge, Case Western Reserve
University, University of Chicago, Drexel University, Fermilab, the
Institute for Advanced Study, the Japan Participation Group, Johns
Hopkins University, the Joint Institute for Nuclear Astrophysics,
the Kavli Institute for Particle Astrophysics and Cosmology, the
Korean Scientist Group, the Chinese Academy of Sciences (LAMOST),
Los Alamos National Laboratory, the Max-Planck-Institute for
Astronomy (MPIA), the Max- Planck-Institute for Astrophysics (MPA),
New Mexico State University, Ohio State University, University of
Pittsburgh, University of Portsmouth, Princeton University, the
United States Naval Observatory, and the University of Washington.

\end{document}